\documentclass[aps,prb,superscriptaddress,twocolumn]{revtex4-2}
\usepackage{graphicx}
\usepackage{dcolumn}
\usepackage{bm}
\usepackage{amssymb}
\usepackage{amsmath}
\usepackage{hyperref}
\usepackage{epsf}
\usepackage{xcolor}
\def\aeti{$\alpha$-(BE\-DT\--TTF)$_2$\-I$_{3}$}
\def\kMn{$\kappa$-(BETS)$_2$\-Mn[N(CN)$_{2}$]$_3$}

\def\kCN{$\kappa$-(BE\-DT\--TTF)$_2$\-Cu$_2$(CN)$_{3}$}
\def\kHgCl{$\kappa$-(BE\-DT\--TTF)$_2$\-Hg\-(SCN)$_{2}$]Cl}
\def\cm{cm$^{-1}$}
\def\TMI{$T_{\rm MI}$}
\draft

\begin{document}
\preprint{kappa-Mn}

\title{Electronic properties of the dimerized organic conductor $\kappa$-(BETS)$_2$Mn[N(CN)$_2$]$_3$}
\author{Marvin Schmidt}
\affiliation{1.~Physikalisches Institut, Universit\"at Stuttgart, Pfaffenwaldring 57,
70569 Stuttgart, Germany}
\author{Savita Priya}
\affiliation{1.~Physikalisches Institut, Universit\"at Stuttgart, Pfaffenwaldring 57,
70569 Stuttgart, Germany}
\author{Zhijie Huang}
\affiliation{1.~Physikalisches Institut, Universit\"at Stuttgart, Pfaffenwaldring 57,
70569 Stuttgart, Germany}
\author{Mark Kartsovnik}
\affiliation{Walther-Meissner-Institut, Bayerische Akademie der Wissenschaften, 85748 Garching, Germany}
 \author{Natalia Kushch}
\affiliation{Institute of Problems of Chemical Physics, Russian Academy of Sciences, 142432 Chernogolovka, Russia}
 \author{Martin Dressel}
\affiliation{1.~Physikalisches Institut, Universit\"at Stuttgart, Pfaffenwaldring 57,
70569 Stuttgart, Germany}

\begin{abstract}
The two-dimensional molecular conductor $\kappa$-(BETS)$_2$Mn[N(CN)$_2$]$_3$ undergoes a sharp metal-to-insulator phase transition at $T_{\rm MI}\approx 21$~K, which has been under scrutiny for many years. We have performed comprehensive infrared investigations along the three crystallographic directions as a function of temperature down to 10~K, complemented by electron spin resonance and dc-transport studies.
The in-plane anisotropy of the optical conductivity is more pronounced than in any other $\kappa$-type BEDT-TTF or related compounds.
The metal-insulator transitions affects the molecular vibrations due to the coupling to the electronic system; in addition we observe a clear splitting of the charge-sensitive vibrational modes
below $T_{\rm MI}$ that evidences the presence of two distinct BETS dimers in this compound.
The Mn[N(CN)$_2$]$_3^-$ layers are determined by the chain structure of the anions resulting in a rather anisotropic behavior and remarkable temperature dependence of the vibronic features.
At low temperatures the ESR properties are affected by the Mn$^{2+}$ ions via $\pi$-$d$-coupling and antiferromagnetic ordering within the $\pi$-spins: The $g$-factor shifts enormously with a pronounced in-plane anisotropy that flips as the temperature decreases; the lines broaden significantly; and the spin susceptibility increases upon cooling with a kink at the phase transition.
\end{abstract}

\date{\today}%
\maketitle

\section{Introduction}
Among the low-dimensional radical cation salts of the BEDT-TTF family, some members containing magnetic ions have sparked broad interest due to the subtle interplay of itinerant $\pi$-electrons in the conducting organic layers and localized $d$-electrons in the insulating anion layers leading to remarkable phenomena such as magnetic order or field-induced superconductivity \cite{Uji01,Balicas01,Uji03}. Here BEDT-TTF stands for bis(ethylene\-di\-thio)tetra\-thia\-fulvalene; its selenium-substituted sibling bis(ethylene\-di\-thio)tetra\-selena\-fulvalene is abbreviated as BETS.
While $\lambda$-(BETS)$_2$FeCl$_4$ and derivatives are certainly the most studied examples among these hybrid salts
\cite{Kobayashi04,Enoki04,Coronado04,Blundell04}, here we focus on the strongly-correlated molecular conductor \kMn, which was first synthesized in 2008 and has drawn attention for rather different reasons ever since
\cite{Kushch08,Zverev10,Vyaselev11b,Vyaselev11a,*Vyaselev12a,*Vyaselev12c,Kartsovnik17,Kushch17,Vyaselev17,Zverev19,Riedl21,Thomas24}.
In contrast to the discrete paramagnetic units Fe$X_4^-$, here the Mn$^{2+}$ ions are connected via diacyanamide bridges.

\begin{figure}
    \centering
        \includegraphics[width=0.9\columnwidth]{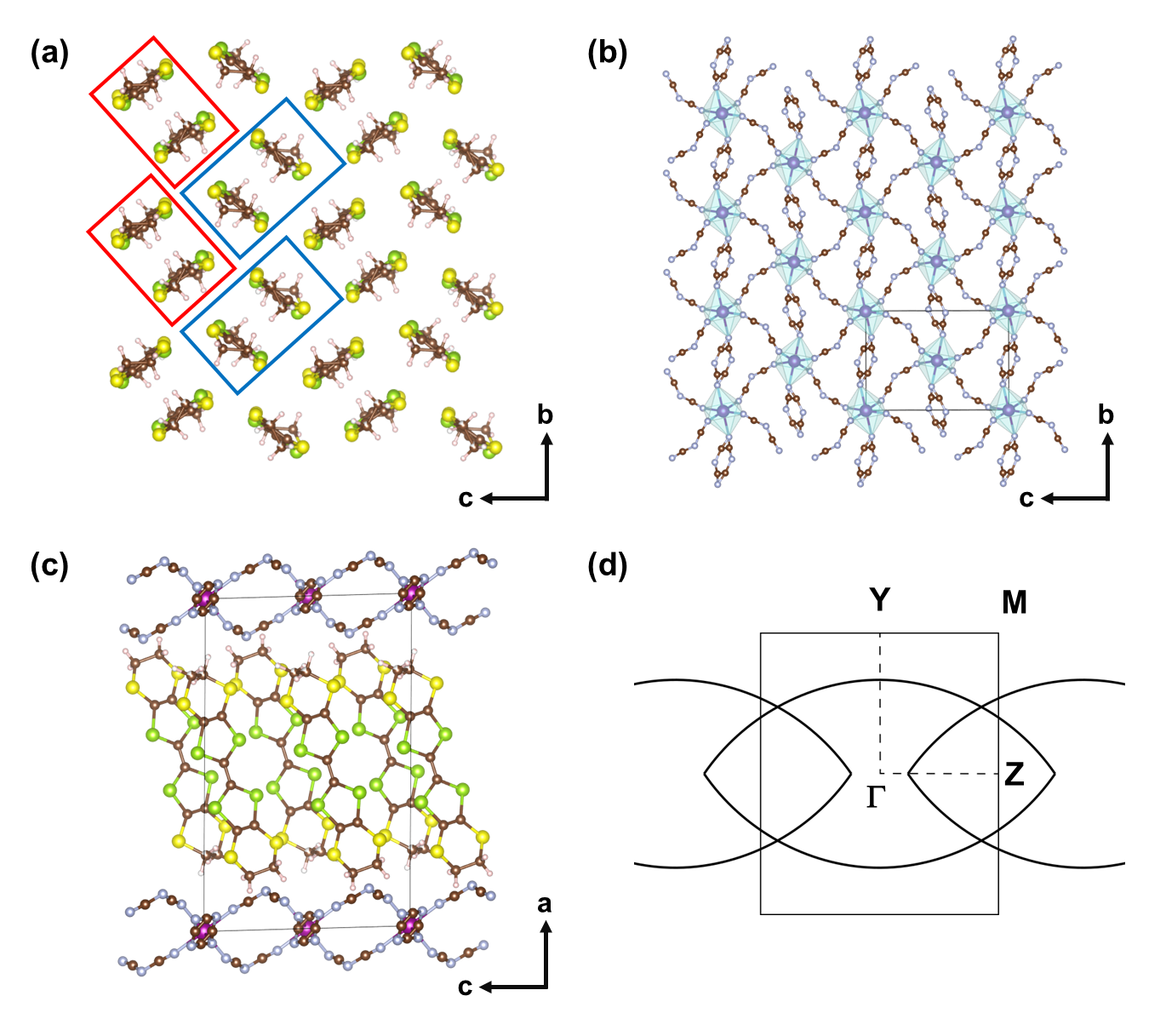}
        \caption{The two-dimensional charge-transfer salt \kMn\ crystallizes in a monoclinic unit cell containing two formula units ($Z=2$); the space group is $P2_1/c$. (a)~Arrangement of the BETS molecules projected onto the $bc$ plane. The distinct dimers A and B are indicated by red and blue; they constitute chains along the $b$-axis. (b) The bulky anions are formed by Mn$^{2+}$ connected to six neighbors via N$\equiv$C-N-C$\equiv$N ligands. There is some structural disorder along the $b$-chains, where both options (up and down) are possible. (c) The layered structure becomes obvious in the $ac$ projection. (d) 2D Fermi surface of \kMn\ calculated according to first-principle by density functional theory (DFT) \cite{Zverev19}.}
    \label{fig:structure}
\end{figure}
According to the well-known $\kappa$-pattern \cite{ToyotaBook,Powell06,*Powell11,Dressel20} the title compound contains (BETS)$_2^+$ dimers that are rotated with respect to each other as depicted in Fig.~\ref{fig:structure}(a).
In contrast to most other $\kappa$-salts, however, two chains are formed along the $b$-axis, where the constituent dimers are distinct by symmetry, labelled A and B. The intrachain coupling is stronger than the interaction between the chains \cite{Zverev10}. This in-plane anisotropy is reflected in the electronic bands, which form a cylindrical Fermi surface crossing the Brillouin boundary along Z-M and leading to a rhombuslike portion around the Z-point with rather flat sides, as displayed in Fig.~\ref{fig:structure}(d) \cite{Zverev10,Zverev19}.
The DFT band-structure agrees well with earlier calculations by the extended Hückel method; one should keep in mind that in both cases correlation effects are neglected.
The importance of electron-electron interaction, however, is unanimously concluded from investigations of the high-field magnetoresistance as well as magnetic quantum oscillations \cite{Zverev10,Kartsovnik17,Zverev19,Oberbauer23}.
When the metallic ground state is stabilized down to lowest temperatures by hydrostatic pressure, Shubnikov–de Haas oscillations are observed for fields above 10~T associated with a classical and a magnetic-breakdown cyclotron orbit on the cylindrical Fermi surface. The enhanced effective cyclotron masses reveal strong renormalization due to sizeable interactions.

From $^1$H- and $^{13}$C-NMR investigations a long-range staggered structure of the $\pi$-electron spins was concluded that occurs already at \TMI\ while the $3d$ Mn$^{2+}$ spin moments form a disordered tilted structure.
Only at low temperatures the $\pi$-$d$-interaction causes antiferromagnetic ordering \cite{Vyaselev12a,*Vyaselev12c,Vyaselev17}. {\it Ab-initio} calculations confirm the rather weak coupling between the two subsystems \cite{Riedl21}.
Torque measurements identify a field-induced transition at temperatures below \TMI\ resembling a spin flop
that is assigned to the spins of $\pi$-electrons localized on the organic BETS molecules \cite{Vyaselev11b}.
Theoretical considerations suggest a spin-vortex crystal order due to ring exchange in \kMn\ \cite{Riedl21}
that might also be relevant to explain the gapped ground state established in the quantum spin liquid candidate
\kCN\ \cite{Riedl19,Miksch21}.

The compound is metallic down to approximately $T_{\rm MI} = 21$~K, when a sharp metal-insulator transition occurs.
Since this is not accompanied by any structural change, a Mott transition due to electron-electron interaction has been proposed \cite{Zverev10} (cf.\ Fig.~\ref{fig:dc}).  The metal-insulator transition can be suppressed by moderate pressure of 0.3~kbar where superconductivity is observed as high as $T_c=5$~K. Initiated by the rhombus-shaped Fermi surface around the Z-point, the possibility of a charge-density-wave formation was discussed \cite{Zverev10}. The degree of nesting, however, might only cause some partial gapping, it certainly is not sufficient to explain the increase in resistivity by several orders of magnitude. There is also no indication of magnetic order that could cause a superstructure.

A closer look at the low-frequency charge dynamics has been obtained by dielectric spectroscopy in combination with
fluctuation spectroscopy \cite{Thomas24}. Below \TMI\ a relaxor ferroelectric behavior is observed that is rather typical for these systems \cite{Tomic15,Lunkenheimer15,Pinteric18}.
Resistance noise spectroscopy reveals fluctuating two-level processes, which are taken as signature of fluctuating polar nanoregions formed already above \TMI.
Upon further cooling these nano-scale polar clusters freeze in a glassy transition, as seen by a drastic slowing down of the charge carrier dynamics.

In order to investigate the electrodynamics of \kMn\ further, we have performed infrared optical measurements for different polarizations and temperatures. In combination with ESR and transport measurements, we obtain the anisotropic charge dynamics, but also elucidate the charge disproportionation that occurs below \TMI. The Mn[N(CN)$_2$]$_3^-$ chains along the $b$-directions are of paramount importance for the electronic properties.

\section{Materials and Methods}
The single crystals of \kMn\ used for our investigations were grown electrochemically \cite{Kushch08} and had the typical size of $20 \times 850 \times 1200~\mu{\rm m}^3$.
The material crystalizes in  $P2_1/c$ space group with $Z=2$, i.e. two dimers per unit cell, A and B.
The  quasi-two-dimensional structure is composed from alternating BETS donor layers,
separated along the $a$-direction by insulating Mn[N(CN)$_2$]$_3^-$ layers, as depicted in Fig.~\ref{fig:structure}(b).
The polymeric anionic structure contains one independent Mn atom located in the inversion center and two independent N(CN)$_2$ ligands. Each Mn atom is linked to six neighboring metal atoms by N(CN)$_2$ bridges. \cite{Kushch08,Schlueter04}. There is some statistical disorder present in the dicyanamide groups located along the $b$-axis. Adjacent chains are coupled with one N(CN)$_2$ bridge above and one below the plane. In contrast to other $\kappa$-phase salts with polymeric anions, here the layers separating the organic molecules are rather spacious.
The BETS molecules compose dimers that are slightly tilted with respect to the $a$-axis [Fig.~\ref{fig:structure}(c)]. We can identify two chains, called A and B, along the $b$-axis formed by independent (BETS)$_2^+$ dimers as illustrated in Fig.~\ref{fig:structure}(a).

The in-plane dc resistivity was measured by standard four-probe technique by attaching four thin gold wires onto the $bc$-surface using carbon paste along the respective orientation. The out-of-plane resistivity was obtained by two contacts on each side. Different crystals exhibit a very similar overall behavior, despite variations in the absolute values.
The infrared properties were probed from close-to-normal incidence optical reflectivity measurements off as-grown crystal surfaces. Besides single crystal measurements in the $a$-direction, we also composed a mosaic of two crystals in order to increase the area. Here the spectral resolution was 1~\cm\ in the far- and mid-infrared region while 4~\cm\ was sufficient at higher frequencies.
Optical spectroscopy was performed at different temperatures ($12~{\rm K}\leq T \leq 300~{\rm K}$) utilizing a Bruker Vertex 80v Fourier transform infrared spectro\-meters with an attached Hyperion IR microscope and a CryoVac Konti Micro helium-flow cryostat.
Due to the limited sample size, reliable data were obtained down to 150~\cm\ and up to 18\,0000~\cm,
employing the most suitable combination of sources, beamsplitters, polarizers, windows, and detectors for each spectral range.
From the merged reflectivity spectra, the optical conductivity was calculated employing  Kramers-Kronig analysis with Hagen-Rubens assumption for the low-frequency range in the metallic regime and extrapolation to a finite reflectivity value below \TMI. For high frequencies we employed Tanner's method \cite{Tanner15} and standard power-law extrapolation.

In addition, the electron spin resonance (ESR) spectra were carried out in a continuous-wave X-band spectrometer (Bruker ESP 300) at 9.5~GHz. The temperature dependence of the ESR properties was measured down to $T=4$~K by utilizing a continuous-flow helium cryostat. These experiments were carried out along different orientations; furthermore the recorded the angular dependence at certain temperatures by rotating the crystals using a goniometer.

\section{Results, Analysis and Discussion}
\subsection{Electronic transport}
\begin{figure}[h]
    \centering
        \includegraphics[width=0.7\columnwidth]{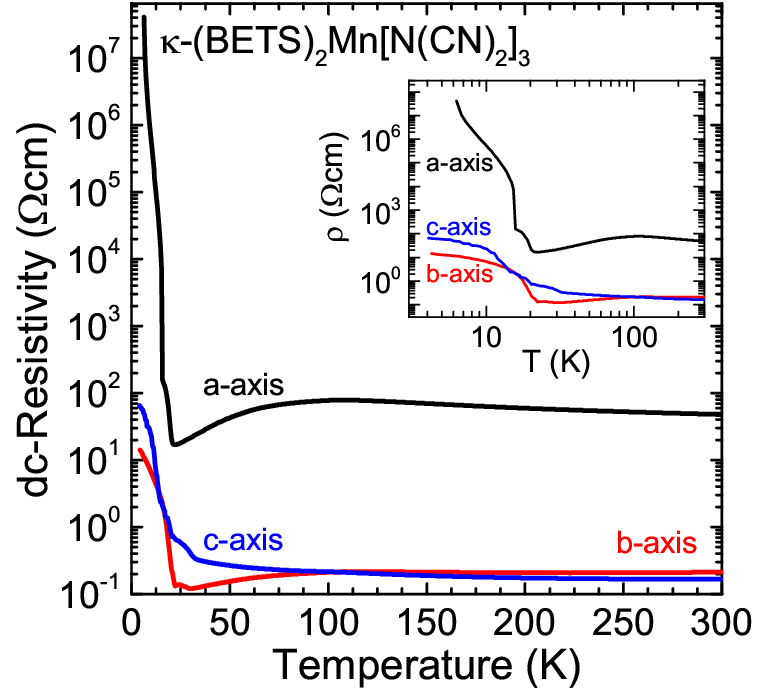}
        \caption{Temperature dependence of the electrical resistivity of \kMn\ measured along different crystallographic axes. In order to magnify the low-temperature behavior the data are displayed on a logarithmic scale in the inset. The metal-insulator transition at $T_{\rm MI}=21$~K is prominent in all three directions.}
    \label{fig:dc}
\end{figure}

Fig.~\ref{fig:dc} displays the dc resistivity of \kMn\ along all three directions.   The room temperature values are $\rho_a =48~\Omega{\rm cm}$, $\rho_b =0.21~\Omega{\rm cm}$ and $\rho_c =0.16~\Omega{\rm cm}$ with a considerable uncertainty in the absolute values
due to the irregular sample geometry.
The overall behavior is similar for the different orientations, and in accord with previous reports \cite{Kushch08}\footnote{In some samples, the minimum in $\rho_c(T)$ is less developed, and occasionally we find a small  increase already around 30~K. Similar observations and sample-to-sample dependences are reported previously \cite{Kushch08,Thomas24}.}.
As the temperature is reduced the resistivity goes through a shallow maximum around 100~K, before a metallic region is entered. The metal-insulator transition occurs at $T_{\rm MI} = 21$~K with a rapid rise in resistivity by several orders of magnitude. The increase in $\rho(T)$ slows down below approximately 16~K, but still remains significant. Indications for a two-step transition are robust: they are  repeatedly observed in different crystals confirming previous reports \cite{Kushch08,Zverev10,Thomas24}.
While the in-plane behavior is very similar at low temperatures, for the out-of-plane direction, the slope $-{\rm d}\rho_a(T)/{\rm d}T$ becomes larger again for $T<5$~K, best seen in the inset of Fig.~\ref{fig:dc}.

\subsection{Optical spectroscopy}
\begin{figure}
    \centering
        \includegraphics[width=1\columnwidth]{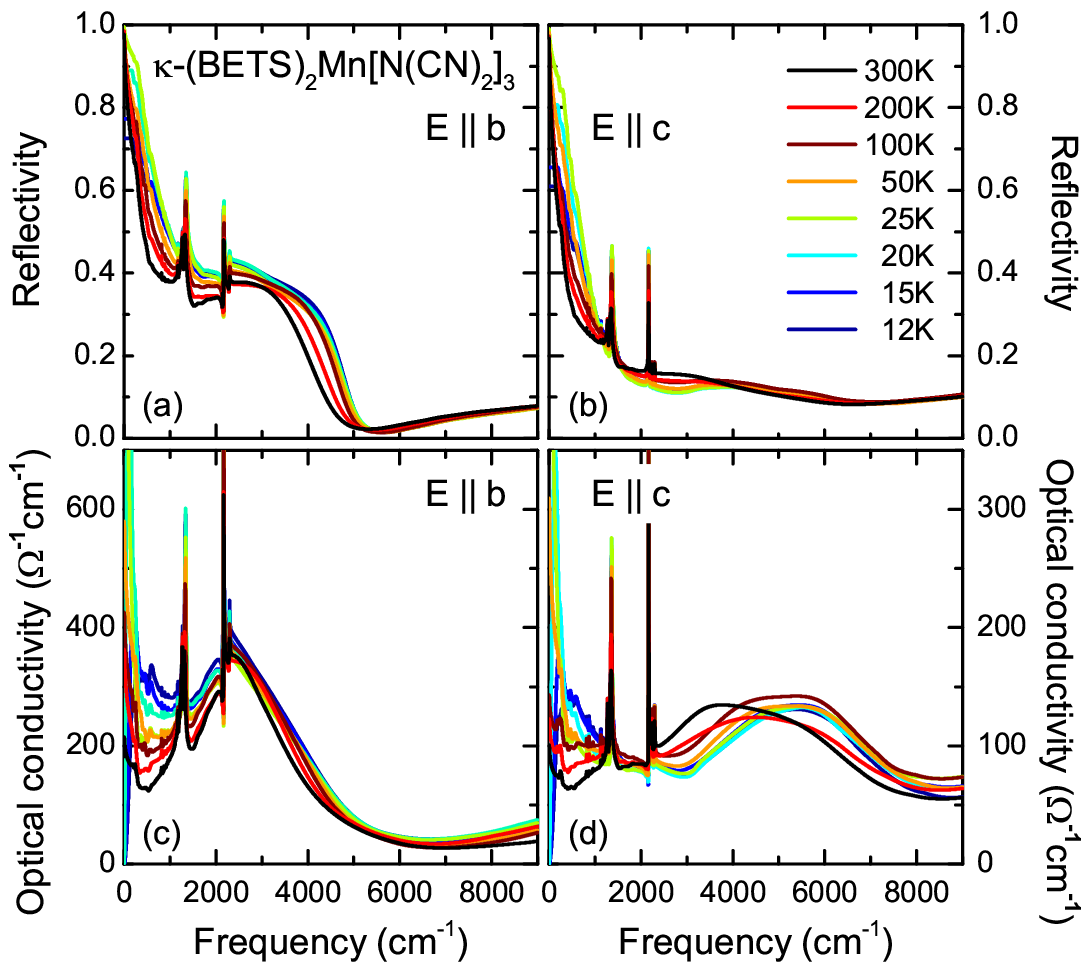}
        \caption{In-plane optical reflectivity of \kMn\ measured for different temperatures as indicated for the polarization (a) $E\parallel b$ and (b) $E\parallel c$. The derived optical conductivity is shown in panels (c) and (d) for the respective directions; note, the conductivity $E\parallel b$ is larger by a factor of 2.}
    \label{fig:optics}
\end{figure}
In Fig.~\ref{fig:optics} the reflectivity and the resulting optical conductivity spectra of \kMn\ are plotted for different temperatures between $T=300$ and 12~K; the light is polarized along the two in-plane directions, i.e.\ $E\parallel b$ and $E \parallel c$. The strong anisotropy of the optical properties can already be seen from the raw data: a pronounced plasma edge is observed in the infrared reflectivity for $E \parallel b$ which is absent in the perpendicular polarization. This results in the rather different conductivity values displayed in Fig.~\ref{fig:optics}(c) and (d). Both spectra are dominated by the large absorption peaks in the mid-infrared:
for the $b$-direction it is centered around $2000$~\cm, while for $E \parallel c$ we find it at $3500$~\cm\ and shifting even higher upon cooling.
These bands correspond to intraband transitions between the lower and upper Mott-Hubbard bands and interband transitions between the dimer bands, respectively and are typical for the $\kappa$-type salts  \cite{Dressel04,Faltermeier07,*Merino08,*Dumm09,Ferber13}.

\begin{figure}
    \centering
        \includegraphics[width=1\columnwidth]{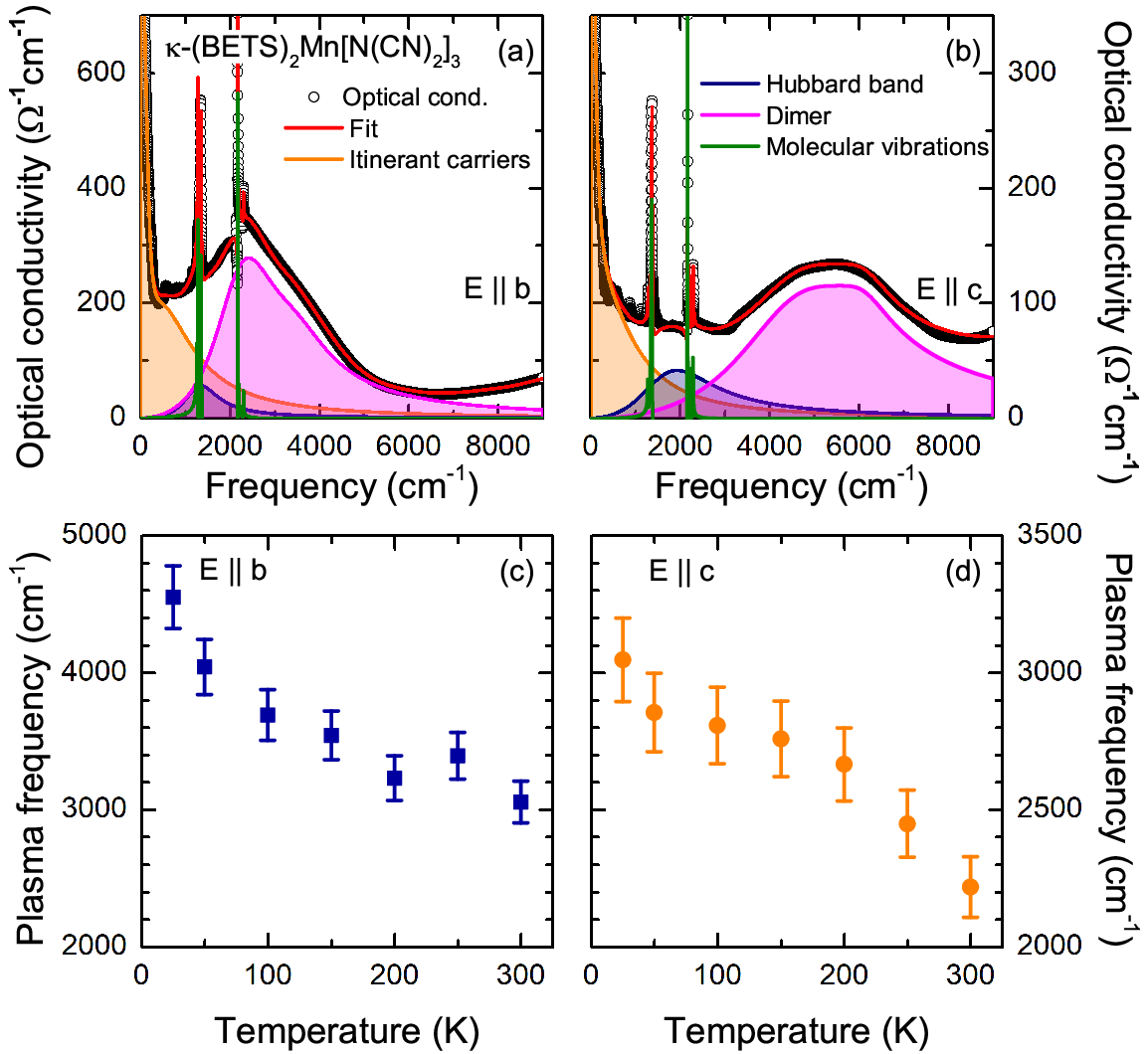}
        \caption{Contributions obtained from Drude-Lorentz fits for the in-plane optical conductivity of \kMn\ at $T = 25$~K for
        the polarizations (a)~$E\parallel b$ and (b)~$E\parallel c$. The plasma frequencies are calculated from the spectral weight of the itinerant carriers by these fits at different temperatures in the metallic regime for (c)~$E\parallel b$ and (d)~$E\parallel c$.}
    \label{fig:Contri}
\end{figure}
Fig.~\ref{fig:Contri} illustrates the different contributions extracted from fitting the optical conductivity by the Drude-Lorentz model \cite{Dressel09}. The sharp molecular vibrations exhibit asymmetric shapes that are best described by Fano functions.
The itinerant carriers at low energy, the mid-infrared absorption peaks and the molecular vibrations are simultaneously fitted with a combination of Drude, Lorentz, and Fano peaks. The plasma frequency $\omega_p$ of the free charge carriers can be calculated from the spectral weight after subtracting the Lorentz and Fano contributions, using
\begin{equation}
{\omega_p^2} = 8 \int_{0}^{\omega_c} \sigma_1(\omega){\rm d}\omega
\end{equation}
where $\sigma_1$ is the real part of the conductivity and $\omega_c$ is the cut-off frequency (10\,000 ~\cm\ in this case) \cite{DresselGruner02}.
With decreasing temperature, the plasma frequency $\omega_p(T)$ of \kMn\ increase for both polarization directions,
reaching around 4500~\cm\ for $E\parallel b$ and 3000~\cm\ for $E\parallel c$ at $T= 25$~K in the metallic region, slightly above the metal-insulator transition. We want to stress the strong in-plane anisotropy, unprecedented  for the $\kappa$-salts.

\subsection{Vibrational modes}
\subsubsection{BETS molecules}
\begin{figure}
    \centering
        \includegraphics[width=0.6\columnwidth]{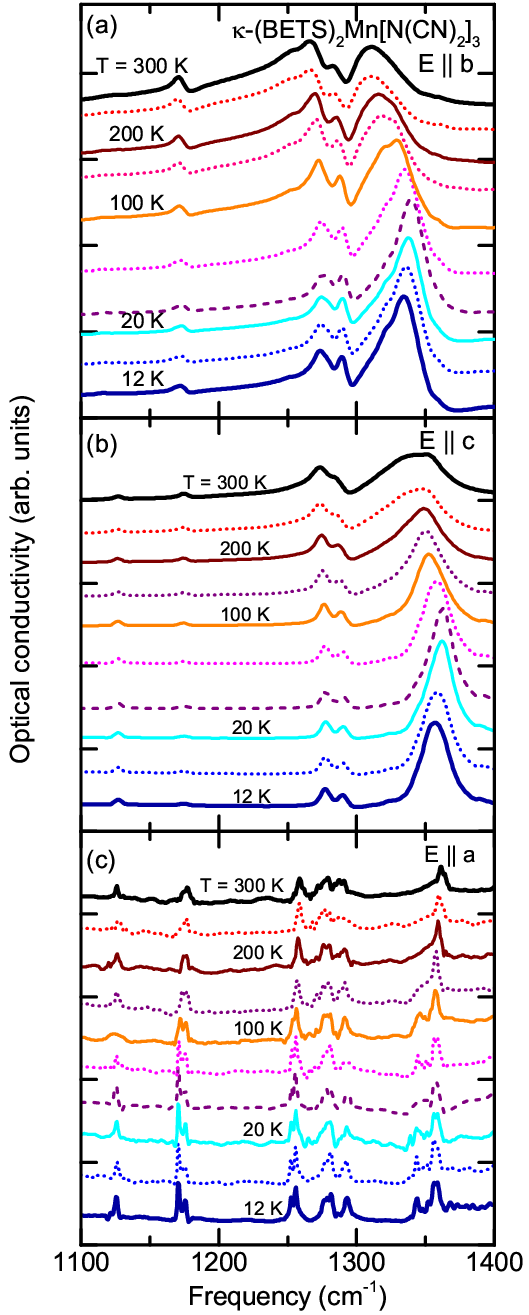}
        \caption{Vibrational modes of \kMn\ in the range from 1100 to 1400~\cm\ for the three polarization directions
        (a)~$E\parallel b$, (b)~$E \parallel c$ and (c) out of plane $E \perp bc$. The optical conductivity spectra measured at the different temperatures ($T=300$, 250, 200, 150, 100, 50, 25, 20, 15, and 12~K from top to bottom) are shifted with respect to each other for clarity reasons.}
    \label{fig:BETS1}
\end{figure}
The characteristic frequency of vibrational modes is extremely sensitive to the variations in the structure and charge distribution;
hence the temperature evolution allows studying configurational changes of crystals.
In the case of organic compounds, molecular vibrations in the mid-infrared are utilized to probe the charge distribution; in particular the C=C bonds of  the BEDT-TTF or BETS molecules are most sensitive fingerprint  \cite{Dressel04,Yamamoto05,Girlando11,*Girlando12,Girlando24}. We should note that only the $\nu_{27}(b_{1u})$ mode is supposed to be infrared active and observed for the out-of-plane polarization, while the $\nu_{2}(a_{g})$ and $\nu_{3}(a_{g})$ modes are fully symmetric and Raman active; it takes electron molecular vibrational (emv) coupling to make them appear in the  in-plane conductivity spectra.

\begin{figure}
    \centering
        \includegraphics[width=0.6\columnwidth]{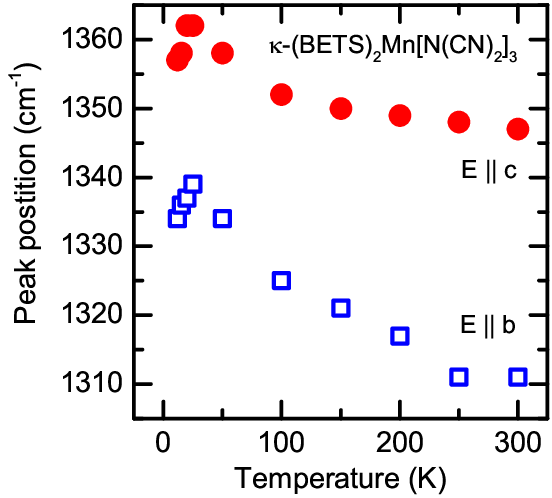}
        \caption{Temperature dependence of the peak position for the fully symmetric $a_g$ modes of the BETS molecule probed with the polarization along the $b$ and $c$ directions. In both cases a pronounced kink is observed at \TMI.}
    \label{fig:peaks1}
\end{figure}

In Fig.~\ref{fig:BETS1} a certain region of the optical conductivity of \kMn\ is plotted
in a waterfall fashion in order to visualize the temperature dependence of these features.  The in-plane optical response is dominated by a broad feature at around 1300~\cm\ that is present for both polarizations. It contains a double peak structure at 1275 and 1290~\cm, which exhibits the usual temperature dependence (hardening and sharpening upon cooling). Possibly it is in antiresonance with the $\nu_5$ mode expected in this range. More spectral weight is found in the band that peaks at 1335 and 1357~\cm\ for $E\parallel b$ as well as for $E \parallel c$, respectively. We ascribe these features to the fully symmetric $a_g$ modes that become infrared active via emv-coupling.
While the bands are rather broad at ambient temperatures,
they significantly narrow with decreasing temperature.
From Fig.~\ref{fig:peaks1} we see that the temperature evolution exceeds the usual hardening, in particular along the $b$ axis. We ascribe the behavior to the structural anisotropy and increase in conductivity as $T$ is reduced.
The maximum values are reached at the transition $T_{\rm MI} = 21$~K.
Note that a pronounced shoulder is present on the low-frequency wing of the $E\parallel b$ peak that we interpret as indication of two separate entities, as we will discuss later.

\begin{figure}
    \centering
        \includegraphics[width=0.6\columnwidth]{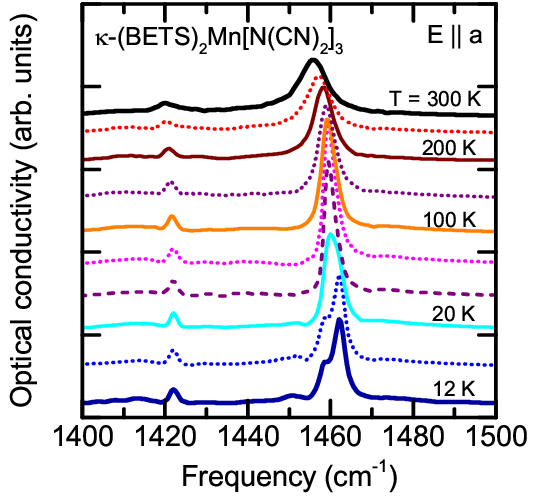}
        \caption{Temperature evolution of the vibrational features of \kMn\ probed perpendicular to the highly conducting $bc$-plane. The dominant vibrational mode $\nu_{27}(b_{1u})$, which is a very sensitive local probe of charge per BETS molecule, splits below $T_{\rm MI}=21$~K. }
    \label{fig:peaks2}
\end{figure}

In the range from 1400 to 1500~\cm\ for $E\parallel a$ we can identify the $\nu_{28}(b_{1u})$ mode around 1420~\cm\ and the more pronounced
$\nu_{27}(b_{1u})$ feature  (cf.\ Fig.~\ref{fig:peaks2}); as the temperature  decreases to \TMI, they exhibit a blue shift by 2 and 4~\cm, respectively.
Most important, however, is the splitting seen in $\nu_{27}(b_{1u})$  at very low temperatures that is a signature of charge disproportionation in \kMn. With only 3~\cm\ the separation is much smaller than observed from typical charge-ordering compounds of the BEDT-TTF family \cite{Dressel04,Girlando11,Tomic15}
Using the standard relation $2\delta_{\rho} = \delta\nu/140~{\rm cm}^{-1}$ \footnote{We note that this relation was derived for BEDT-TTF molecules, but its application can be extended to the BETS molecules under scrutiny here, because the charge sensitive vibrations are basically confined to the C=C bond with the sulfur and selenium almost unaffected, as seen in the supplement of Ref.~\cite{Dressel16}. This is supported by the fact, that the vibrational mode is observed exactly at the same frequency of 1460~\cm\ for the uniform charge of $0.5e$ in both molecues BEDT-TTF and the more heavy BETS.}, we can evaluate the charge imbalance to $2\delta_{\rho} \approx 0.02e$.

From the clearly observed splitting we conclude the presence of two non-equivalent BETS dimers within the unit cell in \kMn, as illustrated in Fig.~\ref{fig:structure}(a), instead of charge disproprtionation between the BETS molecules within a single dimer.
This is strongly supported by very recent first-principle calculations of the infrared and Raman vibrational spectra of $\kappa$-phase salts by Girlando \cite{Girlando24}. Along the long molecular axis,
there exists a strong intramolecular charge oscillation, which results
in a strong interdimer emv coupling, besides the well-known intradimer emv coupling that induces very strong infrared absorptions.

Low-dimensional organic conductors based on TMTTF, BEDT-TTF or related molecules may undergo a charge-order transition
when cooled down. Typically these organic molecules donate uniformly half an electron to the anions at ambient conditions, but in some cases electronic interactions may cause a charge disproportionation below a certain ordering temperature $T_{\rm CO}$ \cite{Seo04,Dressel07,Tomic15}. The charge imbalance $2\delta_{\rho}= \rho_{\rm rich}-\rho_{\rm poor}$ can be up to $0.3e$ as in the case of (TMTTF)$_2$SbF$_6$ and more than $0.6e$ in \aeti\ \cite{Kakiuchi07,Ivek11,Dressel12}.
In dimerized systems, such as $\kappa$-type salts, the coupling between the molecules is typically so strong that the
charge basically remains homogeneous within the dimer
\footnote{There are only a few exceptions, such as
$\kappa$-(BEDT\--TTF)$_4$\-PtCl$_6$$\cdot$C$_6$H$_5$CN, the
triclinic
$\kappa$-(BEDT\--TTF)$_4$\-[$M$(CN)$_6$]\-[N(C$_2$H$_5$)$_4$]$\cdot$3H$_2$O
and the monoclinic
$\kappa$-(BEDT\--TTF)$_4$[$M$(CN)$_6$]\-[N(C$_2$H$_5$)$_4$]$\cdot$2H$_2$O
(with $M$ =  Co$^{\rm III}$, Fe$^{\rm III}$, and Cr$^{\rm III}$)
salts \cite{Doublet94,Maguere96,Swietlik03a,*Swietlik03b,*Swietlik04,Swietlik06,*Ota07,Lapinski13}.
Here the structure is rather complex: the phase transition includes the deformation of the
molecule and the coupling to the anions; accordingly, details of their
physical properties and their electronic states are not well known.},
as monitored by vibrational spectroscopy \cite{Sedlmeier12}.
Recently, the system \kHgCl\ was shown to undergo a charge-order transition at $T_{\rm CO} = 34$~K with a charge imbalance of $2\delta_{\rho}= 0.2e$ \cite{Yasin12,Drichko14,Ivek17,Lohle17,Hassan18,Gati18,Hassan20}; in that particular case the BEDT-TTF molecules are shifted with respect to each other, significantly reducing the intradimer coupling.
Similar to our case, it was suggested \cite{Girlando24} that the charge disproportionation occurs between the dimer chains rather than inside a dimer.

\subsubsection{Anion network}
\begin{figure}
    \centering
        \includegraphics[width=0.6\columnwidth]{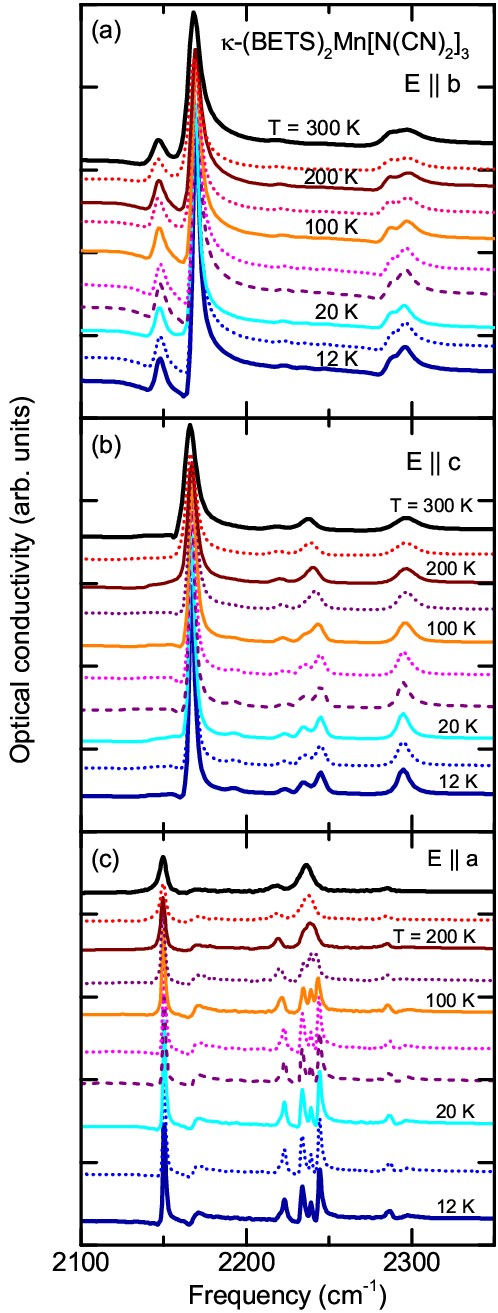}
        \caption{Vibrational modes related to the Mn[N(CN)$_2$]$_3^-$ anions for three polarization directions
        (a)~$E\parallel b$, (b)~$E \parallel c$ and (c) out of plane $E \perp bc$. The optical conductivity spectra measured at the different temperatures ($T=300$, 250, 200, 150, 100, 50, 25, 20, 15, and 12~K) are shifted with respect to each other for clarity reasons.}
    \label{fig:anions1}
\end{figure}

In the present case it is interesting to monitor the vibrational features related to the Mn[N(CN)$_2$]$_3^-$ anionic network, which are most pronounced in the spectral range between 2100 and 2300~\cm\ mainly due to the strong  C$\equiv$N bonds. From Fig.~\ref{fig:anions1} we can identify three groups, from 2150 to 2170~\cm, around 2230~\cm\ and above 2280~\cm; they all disclose an interesting anisotropy and some splitting as the temperature is reduced.

Most prominent are the two features around 2150 and 2170~\cm, which are well pronounced and comparably sharp already at ambient conditions.
For $E \parallel b$ two peaks occur at 2147 and 2168~\cm\ (at room temperature) which shift up only by 1~\cm\ when  cooled down. For the $c$-polarization only the interchain links are probed, resulting in a single strong mode. For the out-of-plane polarization, these vibrations are basically absent but the peak at 2149~\cm\ appears very sharp. No splitting is observed in these vibrational features, but all of them are characterized by an asymmetric shape due to some coupling.

\begin{figure}
    \centering
        \includegraphics[width=0.93\columnwidth]{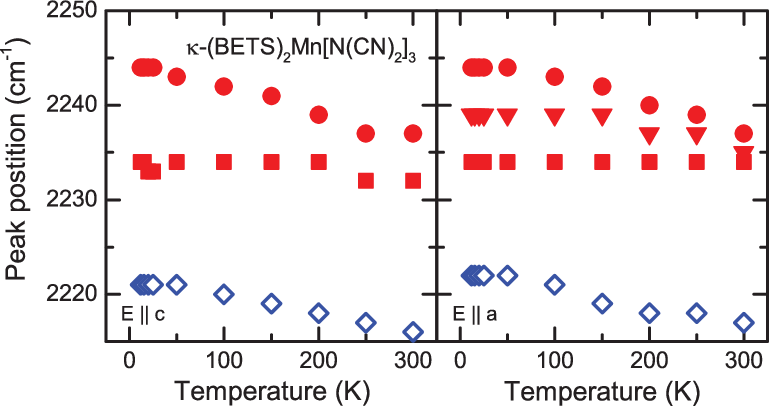}
        \caption{Temperature evolution of the peaks that constitute the 2200 to 2235~\cm\ bands in the $E\parallel c$ and $E\parallel a$ polarization. While at $T=300$~K mainly two maxima are identified (red and blue symbols), the higher frequency feature splits into three lines as the temperature is reduced. Only for $E\parallel a$ three sub-peaks can be identified unambiguously.}
    \label{fig:anions2}
\end{figure}
The modes grouped between 2220 and 2250~\cm, however, exhibit a remarkable temperature dependence.
While the band at 2297~\cm\ is isolated for $E\parallel c$, it comes with a shoulder about 10~\cm\ lower in frequency for the $b$-direction. Small indications of this vibration can be identified also for the perpendicular direction. Very interesting are the modes at 2236 and 2218~\cm\ that are most prominent for the out-of-plane polarization.
The lower peak sharpens when the temperature is reduced and moves up to 2223~\cm, with some saturation observed below 50~K.
The
stronger high-frequency modes separate into three: 2234, 2239 and 2244~\cm, best seen for  $T <200$~K.
As illustrated in Fig.~\ref{fig:anions2} this splitting occurs gradually, with no pronounced kink, which confirms the absence of any symmetry change.
Instead, the octahedral Mn configuration is subject to a Jahn-Teller distortion resulting in inequivalent N(CN)$_2$ ligands which become visible at low temperatures. The orientation of the Jahn-Teller distortion is known to affect the magnetic interaction between Mn$^{\rm II}$-complexes \cite{Alowasheeir18,Bikas19}.
It is important to note, however, that no pronounced modifications of these vibrational features are related with the metal insulator transition at \TMI. We conclude that the magnetic interaction among the Mn-ions does not change significantly at that temperature.

The high-frequency modes below 2300~\cm\ resembles the low-frequency group at 2150~\cm: for $E\parallel b$ it exhibits a double structure, while for the two other polarization directions, one of the vibrations is dominant, respectively.

Fig.~\ref{fig:structure}(b) clearly illustrates why the anionic structure exhibits such an anisotropic behavior.
The Mn-N(CN)$_2$ arrangement along the $b$-direction with some structural disorder is rather robust upon cooling. As a result, the vibrational features vary only slightly. Between two adjacent $b$-chains, there is only one type of contact (alternating above and below the plane) resulting in only one strong vibrational mode.

\subsection{Magnetic spectroscopy}
\begin{figure}
    \centering
        \includegraphics[width=0.65\columnwidth]{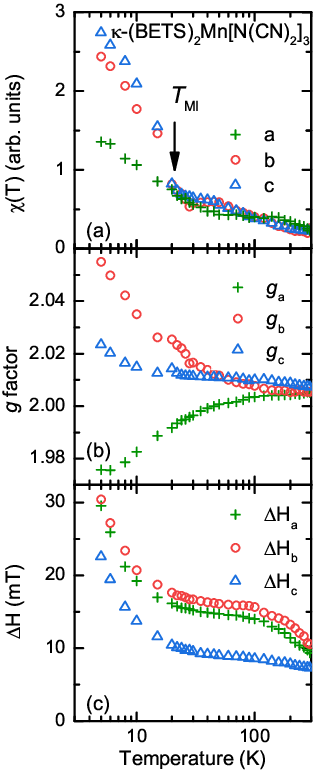}
        \caption{ESR parameters extracted from temperature-dependent X-band measurements on \kMn\ along the three crystallographic directions $H \parallel a$, $b$, and $c$. Note the logarithmic temperature scale. (a)~The spin susceptibility is proportional to the integrated absorption intensity. While the experiments within the $bc$-plane are measured on the same crystal, the $a$-axis data are taken on a different crystal and had to be normalized by a factor of 3. (b)~The $g$-factor starts to deviate at lower temperatures due to the strong effect of the manganese ions. (c)~Temperature behavior of the ESR linewidth $\Delta H$.}
    \label{fig:ESR1}
\end{figure}

The temperature dependence of the ESR findings for \kMn\ obtained by X-band spectroscopy within the $bc$-plane and perpendicular to it is plotted in Fig.~\ref{fig:ESR1}. At high temperatures the finite conductivity causes a slightly asymmetric lineshape (Dysonian), which becomes a simple Lorentzian below approximately 26~K, in good agreement with NMR measurements \cite{Vyaselev11a,Vyaselev11b}.
In all directions the spin susceptibility continuously increases as the temperature is reduced; below approximately 25~K it rapidly rises, marking the transition at \TMI. While the present experiments are carried out at rather low fields of approximately 0.3~T, this phase transition was not detected by previous susceptibility measurements using a superconducting quantum interference device (SQUID) magnetometer operated at 7~T \cite{Kushch08}\footnote{The magnetization measured by a SQUID is mostly determined by Mn$^{2+}$ spins and is not very sensitive to the $\pi$-electron spins.}.

The $g$-factor is a measure of the local magnetic field and in Fig.~\ref{fig:ESR1}(b) it is plotted as a function of temperature for the external field aligned along the three crystal directions. Below $T\approx 100$~K magnetic interactions among the $\pi$-spins and effects of the Mn$^{2+}$ ions become obvious as the signal strongly deviates from the free electron value.
We find that $g(T)$ gradually decreases for the external field oriented out of plane, but strongly rises for the in-plane direction. This becomes even more significant when the transition is approached. For $H\parallel b$ the effect is most pronounced.
The linewidth $\Delta H$ increases as the temperature is lowered [cf.\ panel (c)]; it is smallest for the $c$ direction. Again below 25~K the arranged magnetic moments cause a significant local field and enhanced scattering of the $\pi$-electron spins leads to a rapid rise of $\Delta H(T)$. Proton NMR spectroscopy came to the same conclusion \cite{Vyaselev11b}.

Compared to other $\kappa$-salts with non-magnetic metal ions in the anion layer, such as
the spin liquid candidates \kCN\ or $\kappa$-(BE\-DT\--TTF)$_2$\-Ag$_2$(CN)$_{3}$, the spin susceptibility and the linewidth of \kMn\ become much larger at low temperatures, evidencing the strong coupling among the magnetic moments. This contribution is negligible at elevated temperatures. Nevertheless, despite the sizeable $\pi$-$d$-interaction, our X-band measurements cannot resolve the hyperfine structure of the $S=5/2$ Mn$^{2+}$; it just contributes to the broadening of the lines at low temperatures in the insulating state \footnote{We suggest Q-band and W-band ESR measurements for a better resolution in order to gain more insight; additionally the hyperfine couplings could be studied applying dedicated spectroscopies, such as ENDOR or HYSCORE. }.

\begin{figure}
    \centering
        \includegraphics[width=1\columnwidth]{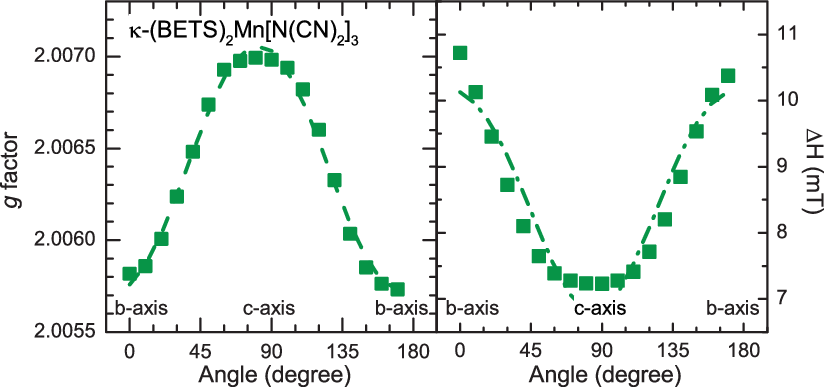}
        \caption{Angular dependence of the room-temperature ESR parameters upon rotating in the $bc$-plane of \kMn. The $g$-factor and the linewidth $\Delta H$ follow a simple cosine behavior, where the $g$-value is maximum along the $c$-direction, while the line is broader for $H\parallel b$.}
    \label{fig:ESR2}
\end{figure}

For the quasi-one-dimensional salts, which undergo a charge-order transition,
such as (TMTTF)$_2$SbF$_6$, a remarkable anomaly of the ESR linewidth was observed at low temperatures \cite{Yasin12b}: Due to anisotropic Zeeman interaction the linewidth doubles its periodicity when rotated in the plane. In the present case of \kMn\ no such doubling could be resolved. The angular dependence of the $g$-factor and linewidth $\Delta H$ displayed in Fig.~\ref{fig:ESR2} follows a simple cosine behavior; most importantly, this does not change when cooling down to 2~K. We conclude that the coupling of the BETS dimers A and B is not sufficient to change the symmetry in the ESR response, although they carry a slightly different charge.
Also rotation in the $ab$-plane yields the regular cosine dependence of $g$-value and linewidth $\Delta H$ at low temperatures, when the coupling to the Mn$^{2+}$ spins dominates. At elevated temperatures the response becomes more complex since the different contributions become comparable.

\section{Conclusions}

Compared to other $\kappa$-salts, the electronic response  of \kMn\ single crystals appears very anisotropic in the conducting plane,
in particular as far as the infrared  properties are concerned. Chains of BETS-dimers arranged along the $b$-axis dominate the
electronic coupling.
The origin of the metal-insulator transition observed at $T_{\rm MI} = 21$~K cannot be completely resolved.
Our ESR measurements prove the importance of magnetic coupling at low temperatures inferring a spin-density-wave transition. However, nesting is possible only in small parts of the Fermi surface, making instabilities rather unlikely. Electronic correlations certainly play a role, but the compound seems not be a simple Mott insulator.

From our optical and magnetic studies we conclude that charge disproportionation is observed
in the title compound and appears differently compared to the established charge-ordered insulators. This charge imbalance could be arising as a consequence of a phase transition instead of driving it, mainly illustrated by Fig.~\ref{fig:peaks2}: there are no well separated peaks detected as in the case of \aeti\ or \kHgCl, but only a distinct shoulder, relating to a charge disproportionation of $2\delta_{\rho}= 0.02e$. Due to the strong coupling between the dimer molecules of this $\kappa$-phase organic conductor, such a minor imbalance within the (BETS)$_2$ entity could not be identified; instead it leads to a broadening of the charge-sensitive vibrational feature.
The comparison of the measured and calculated vibrational spectra clearly shows that the intradimer inversion center symmetry is not broken, i.e.\ the charge imbalance does not reside on the dimers. Instead differently charged dimers become energetically favorable at low temperatures. This is basically what we observe in \kMn, suggesting stripes of charge-rich and charge-poor dimers in the insulating phase.
This effect, however, does not lead to a doubling of the periodicity in the ESR linewidth, because the charge imbalance is not large enough and the antiferromagnetic order is overwhelming at low temperatures. Pressure- and field-dependent optical studies might help to resolve the puzzle of \kMn.

\acknowledgments

We acknowledge the technical supports from Gabriele Untereiner and Sudip Pal.
The project was supported by the Deutsche Forschungsgemeinschaft (DFG) via DR228/39-3.

%

\end{document}